\documentclass[pre,twocolumn,aps,floatfix,superscriptaddress]{revtex4}
\usepackage{amsmath}
\usepackage{amsfonts}
\usepackage{amssymb}
\usepackage{graphicx}
\usepackage{ucs}
\usepackage[utf8x]{inputenc}
\usepackage{upgreek}
\usepackage{subfigure}

\renewcommand{\Re}{\operatorname{Re}}
\renewcommand{\Im}{\operatorname{Im}}
\newcommand{\BigO}[1]{\ensuremath{\operatorname{\mathcal{O}}\bigl(#1\bigr)}}
\renewcommand{\vec}[1]{\mathbf{#1}}

\begin{document}

\title{Long wavelength properties of phase field crystal models with second order dynamics}

\author{V. Heinonen}
\affiliation{COMP Centre of Excellence,
Department of Applied Physics,
Aalto University, School of Science,
P.O.Box 11100,
FI-00076 Aalto
Finland}
\email{vili.heinonen@aalto.fi}
\author{C. V. Achim}
\affiliation{COMP Centre of Excellence,
Department of Applied Physics,
Aalto University, School of Science,
P.O.Box 11100,
FI-00076 Aalto
Finland}
\author{T. Ala-Nissila}
\affiliation{COMP Centre of Excellence,
Department of Applied Physics,
Aalto University, School of Science,
P.O.Box 11100,
FI-00076 Aalto
Finland}
\affiliation{Department of Physics, Brown University, 
Providence RI 02912-1843, USA}

\begin{abstract}
The phase field crystal (PFC) approach extends the notion of phase field models by describing the topology of the microscopic structure of a crystalline material. One of the consequences is that local variation of the interatomic distance creates an elastic excitation. The dynamics of these excitations poses a challenge: pure diffusive dynamics cannot describe relaxation of elastic stresses that happen through phonon emission. To this end, several different models with \emph{fast dynamics} have been proposed. In this article we use the amplitude expansion of the PFC model to compare the recently proposed hydrodynamic PFC amplitude model with two simpler models with fast dynamics. We compare these different models analytically and numerically. The results suggest that in order to have proper relaxation of elastic excitations, the full hydrodynamical description of the PFC amplitudes is required. 
\end{abstract}

\maketitle

\section{Introduction}
The time evolution of the microstructure of crystalline materials is mostly governed by diffusive slow phenomena. This makes materials modeling at atomistic length and time scales a demanding task. Traditional atomistic methods such as molecular dynamics have to solve the time evolution equations at the time scale of atomistic vibrations making it extremely difficult to extend the simulation time scales to diffusive, entropy driven phenomena. One attempt to overcome this limitation come with the introduction of the field theoretical framework of phase field crystal (PFC) models proposed by Elder \textit{et al.} \cite{Elder:2002eq,Elder:2004}.

The main idea behind the PFC approach is that the positions of the atoms are given by an atomic number distribution function that is phenomenologically connected to the canonical distribution function of the microscopic structure. The advantage of this approach is that the positions of the atoms are taken to be thermally averaged and no need for solving  the fast, atomistic time scales, remain. The PFC approach has been hugely successful in describing a wide range of static and dynamical material properties \cite{Emmerich:2012ko}.

One of the advantages of the PFC model is the intrinsic incorporation of elastic excitations. The PFC equations introduce an interatomic length scale that can be varied locally giving rise to elastic excitations. This poses a challenge for the dynamics since elastic excitations should create vibrations of the lattice that cannot be described using diffusive first order dynamics. The first attempt to tackle this problem was with the introduction of the modified PFC (MPFC) model in which a second order time derivative is added in the time evolution equation of the system \cite{Stefanovic:2006fg,Galenko:2009hr}. The MPFC model is appealing due to its simplicity and is able to introduce another time scale in the dynamics allowing for faster relaxation of elastic excitations. However, the MPFC remains hard to motivate physically and fails in describing phonon modes \cite{Majaniemi:2007iy,Majaniemi:2008cq}.

In order to describe the lattice vibrations more realistically, PFC equations have been coupled to the time evolution equation of a hydrodynamical momentum density \cite{Baskaran:2014ko}. Two main problems arise with this approach. First, the oscillating nature of the PFC solid creates spurious flows at interatomic length scale. Second, it is hard to come up with a realistic way to incorporate dissipation in the momentum density equation since macroscopic equations such as the Navier-Stokes equation consider smooth fields and it is not clear how to extend the dissipation to atomistic length scales. These problems suggest the need for a coarse graining procedure for the momentum density field. 

In Ref.~\cite{Toth:2013ie} the authors used special Fourier filters to coarse grain the velocity and the density fields that are then coupled to the microscopic density field. Another approach for smoothing out the velocity field was introduced in \cite{Praetorius:2015co}. The authors consider a colloidal solution assuming an extreme viscosity for the colloidal particles. This way the full hydrodynamical equations are solved essentially only in the solution.

The most recent approach for coupling the momentum density with the microscopic system described in Ref.~\cite{Heinonen:2015} uses the amplitude expansion framework of the PFC model introduced by Goldenfeld \textit{et al.} \cite{Goldenfeld:2005dq,Athreya:2006hs}. The amplitude expansion of the PFC model takes advantage of the fact that the solution for the PFC density is close to a one-mode approximation of a given crystal symmetry. Instead of solving the PFC density, the amplitude equations consider the envelope of the periodically varying PFC density. This envelope is slowly varying in space making it suitable for coupling to a slowly varying velocity field.

In this article we study the difference of the amplitude description of the MPFC model and the recently proposed hydrodynamical amplitude expansion model. These are the simplest PFC models with fast dynamics. The article is organized as follows: Sec.~\ref{sec:background} gives background on the PFC model and modified PFC dynamics. Amplitude expansion of the PFC model is shortly discussed and the different dynamical schemes are introduced in Sec.~\ref{sec:amplitude_expansion}. Phonon spectrum and small deformations are studied analytically in Sec.~\ref{sec:small_displacement_limit}. The different schemes are compared numerically in Sec.~\ref{sec:numerical_results} and finally the results are summarized and we conclude in Sec.~\ref{sec:summary_and_conclusion}.

\section{Background}
\label{sec:background}
Here we motivate the simple fast dynamics by considering linearized hydrodynamics. Let the free energy of the PFC system $F_{\text{PFC}}[n]$ be defined through PFC density $n$. We can write down a momentum density continuity equation with the help of continuity equation for $n$
\begin{equation}
\partial_t n = - \nabla \cdot (n \vec{v}),
\label{eq:PFC_mass_conservation}
\end{equation}
as
\begin{equation}
n (\partial_t \vec{v} + \vec{v}\cdot \nabla \vec{v})= - n \nabla \frac{\delta F_{\text{PFC}}}{\delta n} + \mathcal{D}(n) \vec{v}.
\label{eq:PFC_velocity_conservation}
\end{equation}
Here $\delta F / \delta n$ is the chemical potential that acts as a source term for the momentum density $n \vec{v}$ and $\mathcal{D}$ is some operator that defines dissipation. Assuming that velocity and its derivatives are small we can expand this equation up to the first order in $\vec{v}$ giving
\begin{equation}
n \partial_t \vec{v}  = - n \nabla \frac{\delta F_{\text{PFC}}}{\delta n} + \mathcal{D} \vec{v}.
\label{eq:linearisedPFCvelocity}
\end{equation}
Taking time derivative of Eq.~\eqref{eq:PFC_mass_conservation} gives
\begin{equation*}
\partial_t^2 n = - \nabla \cdot (\vec{v} \partial_t n + n \partial_t \vec{v}),
\end{equation*}
from which we get 
\begin{equation*}
\begin{split}
\partial_t^2 n &= \nabla \cdot \left[ 
\vec{v} \nabla \cdot (n \vec{v}) + n \nabla \frac{\delta F_{\text{PFC}}}{\delta n}  - \mathcal{D} \vec{v}
\right] \\
& \approx \nabla \cdot \left[ 
 n \nabla \frac{\delta F_{\text{PFC}}}{\delta n}  - \mathcal{D} \vec{v}
\right]
\end{split}
\end{equation*}
with the help of Eqs.~\eqref{eq:PFC_mass_conservation} and \eqref{eq:linearisedPFCvelocity}. Here we have discarded nonlinear terms in $\vec{v}$. To continue with the calculation we need to define the dissipation term $\mathcal{D}$. For Langevin dissipation we have $\mathcal{D}=-\alpha n$ giving
\begin{equation*}
\partial_t^2 n  = \nabla \cdot \left( 
n \nabla \frac{\delta F_{\text{PFC}}}{\delta n} 
\right)
+ \alpha \nabla \cdot (n \vec{v}).
\end{equation*}
The latter part is given by Eq.~\eqref{eq:PFC_mass_conservation} and finally the time evolution can be written in terms of $n$ as
\begin{equation}
\partial_t^2 n  + \alpha \partial_t n = \nabla \cdot \left( 
n \nabla \frac{\delta F_{\text{PFC}}}{\delta n} 
\right).
\label{eq:fullMPFC}
\end{equation}
Often we simplify the source term by replacing the right hand side of Eq.~\eqref{eq:fullMPFC} by $\nabla^2 (\delta F_{\text{PFC}} / \delta n)$ giving
\begin{equation}
\partial_t^2 n  + \alpha \partial_t n = \nabla^2 \frac{\delta F_{\text{PFC}}}{\delta n} .
\label{eq:MPFC}
\end{equation}

A system with dynamics described by Eq.~\eqref{eq:MPFC} is known as the modified phase field crystal model (MPFC) \cite{Stefanovic:2006fg}. Here the dissipation is controlled through the $\alpha$ parameter. We get the regular PFC model by taking the large $\alpha$ limit given by
\begin{equation}
\partial_t n = \nabla^2 \frac{\delta F_{\text{PFC}}}{\delta n} .
\end{equation}

\section{Amplitude expansion}
\label{sec:amplitude_expansion}

In this article we are interested in the long-wavelength behaviour of the PFC models and therefore we use the amplitude expansion framework \cite{Goldenfeld:2005dq,Athreya:2006hs}. In this framework the PFC density $n$ is approximated by its one-mode approximation 
\begin{equation}
n \approx \rho + \sum_j  \left(
\eta_j e^{i \vec{q}_j \cdot \vec{r}} + \text{C.C.}
\right),
\label{eq:field_approximation}
\end{equation}
where only the reciprocal lattice vectors $\vec{q}_j$ of the first star are taken into account. We choose a representation for a 2D hexagonal lattice as $\vec{q}_j$ as $\vec{q}_1=(-\sqrt{3}/2,-1/2)$, $\vec{q}_2=(0,1)$ and $\vec{q}_3=(\sqrt{3}/2,-1/2)$. Note that $|\vec{q}_j|=1$. For details of the coarse-graining procedure, see Ref.~\cite{ Athreya:2006hs}.

The amplitudes $\eta_j$ are taken to be complex to allow for displacements. Consider a change of amplitudes $\eta_j \to  \eta_j \exp{[-i\vec{q}_j \cdot \vec{u}(\vec{r})]}$ with some field $\vec{u}$. The approximation for the microscopic field changes as 
\begin{equation*}
\begin{split}
n &\approx \rho + \sum_j  \left(
\eta_j e^{-i \vec{q}_j \cdot \vec{u} } e^{i \vec{q}_j \cdot \vec{r}} + \text{C.C.}
\right) \\
&= \rho + \sum_j  \left(
\eta_j  e^{i \vec{q}_j \cdot (\vec{r}- \vec{u})} + \text{C.C.}
\right) .
\end{split}
\end{equation*}
This shows that the field $\vec{u}$ is a displacement field. 

We will discuss five different models with the same free energy 
\begin{equation}
\begin{split}
F = &\int d\vec{r} \left\lbrace \vphantom{\sum_{j=1}^3}
\frac{B^{\ell}}{2} \rho^2 -\frac{\tau}{3} \rho^3
+ \frac{\nu}{4} \rho^4 
+ \frac{\tilde{B}^x}{2}|\nabla \rho|^2  \right. \\ &\left.
+ \left( \frac{\Delta B}{2} - \tau \rho  +\frac{3\nu}{2} \rho^2 \right) A^2
+ \sum_{j=1}^3 B^x |\mathcal{G}_j \eta_j|^2  \right. \\ &\left.
+ \left( 6 \nu \rho -2\tau \right)  \left( \prod_{j=1}^3 \eta_j + \textrm{C.C.} \right) 
+\frac{3\nu}{4} A^4  \right. \\ &\left.
- \frac{3\nu}{2} \sum_{j=1}^3 |\eta_j|^4
\right\rbrace,
\end{split}
\label{eq:energy}
\end{equation}
where $\mathcal{P} = B^{\ell} - \tilde{B}^x \nabla^2$, $\mathcal{G}_j = (\nabla^2 + 2 i \vec{q}_j\cdot \nabla)$,
$A^2 = 2 \sum_{j=1}^3 |\eta_j|^2$, and C.C. denotes the complex conjugate. These models will be described in detail in the following sections.


\subsection{MPFC amplitude expansion (MPFCA) model }
The MPFCA model is the amplitude expansion of Eq.~\eqref{eq:MPFC}. The dynamics are described by
\begin{equation}
\partial_t^2 \eta_j + \alpha \partial_t \eta_j = - \frac{\delta F}{\delta \eta_j^*}.
\label{eq:MPFCAevolution}
\end{equation}
Again the dissipation happens through the term with the parameter $\alpha$. The density in Eq.~\eqref{eq:energy} is taken to be constant.

\subsection{Augmented MPFC amplitude expansion (AMPFCA) model}
For this model we choose a different type of dissipation term. The dynamics are described by
\begin{equation}
\partial_t^2 \eta_j - \alpha \mathcal{Q}_j^2 \partial_t \eta_j = - \frac{\delta F}{\delta \eta_j^*},
\label{eq:AMPFCAevolution}
\end{equation}
where $\mathcal{Q}_j^2 = (\nabla + i \vec{q}_j)^2 = \nabla^2 + 2i \vec{q}_j \cdot \nabla - 1$.

This model can be derived from a PFC equation similar to Eq.~\eqref{eq:fullMPFC} where $\alpha$ has been replaced with $-\alpha \nabla^2$. The amplitude representation of the Laplacian $\nabla^2$ is $\mathcal{Q}_j^2$.

This model can be examined through energetics. Let us define a kinetic energy
\begin{equation}
\begin{split}
T[\eta_j, \eta_j^*] &= \int d\vec{r} \left[
\sum_{j=1}^3 \left|
\partial_t \eta_j
\right|^2
\right] \\
&= \int d\vec{r} 
\left[
\sum_{j=1}^3 
\partial_t \eta_j \partial_t \eta_j^*
\right] .
\end{split}
\end{equation}
We define a total effective Hamiltonian $\mathcal{H}=T+F$. The time-evolution of the total energy becomes
\begin{equation*}
\begin{split}
\partial_t \mathcal{H} &= \partial_t T + \partial_t F \\
&=
\int d\vec{r} \left[
\sum_{j=1}^3 \left(
\partial_t \eta_j^* \partial_t^2 \eta_j + \text{C.C.}
\right)
\right] \\
&+ \int d\vec{r} \left[
\sum_{j=1}^3 \left(
\partial_t \eta_j^* \frac{\delta F}{\delta \eta_j^*} + \text{C.C.}
\right)
\right] \\
&=
\int d\vec{r} \left[
\sum_{j=1}^3 \left(
\partial_t \eta_j^* \partial_t^2 \eta_j 
+ \partial_t \eta_j^* \frac{\delta F}{\delta \eta_j^*} 
\right)
+ \text{C.C.}
\right] \\
&=
\int d\vec{r} \left\lbrace
\sum_{j=1}^3 \left[
\partial_t \eta_j^* \left( \partial_t^2 \eta_j 
+  \frac{\delta F}{\delta \eta_j^*} \right)
\right]
+ \text{C.C.}
\right\rbrace.
\end{split}
\end{equation*}
Inserting  $\partial_t^2 \eta_j + \delta F / \delta \eta_j^*$ using Eq.~\eqref{eq:AMPFCAevolution} gives
\begin{equation*}
\begin{split}
\partial_t \mathcal{H}
&= 
\int d\vec{r} \left\lbrace
\sum_{j=1}^3 \left[
 \partial_t \eta_j^* \cdot \alpha \mathcal{Q}_j^2 \partial_t \eta_j 
\right]
+ \text{C.C.}
\right\rbrace \\
&= 
-\alpha \int d\vec{r} \left\lbrace
\sum_{j=1}^3  \left[
 (\mathcal{Q}_j^*  \partial_t \eta_j^*) \cdot (\mathcal{Q}_j \partial_t \eta_j) 
\right]
+ \text{C.C.}
\right\rbrace. 
\end{split}
\end{equation*}
Here we have used integration by parts. Finally we get
\begin{equation}
\partial_t \mathcal{H} = -2 \alpha \int d\vec{r} \left\lbrace
\sum_{j=1}^3  
 |\mathcal{Q}_j  \partial_t \eta_j|^2
\right\rbrace \leq 0. 
\end{equation}
This shows how the parameter $\alpha$ controls dissipation. A similar result can be obtained for the MPFCA model with the substitution $\mathcal{Q}_j^2 \to -1$.

\subsection{PFC amplitude expansion with hydrodynamics (HPFCA) }
This model is described in depth in Ref.~\cite{Heinonen:2015}. The dynamics are given by
\begin{equation}
\frac{D \vec{v}}{D t} = 
- \nabla \frac{\delta F}{\delta \rho} 
- \frac{1}{\rho} \sum_{j=1}^3 
\left[  \eta_j^* \mathcal{Q}_j 
\frac{\delta F}{\delta \eta_j^*}  
 + \textrm{C.C.}
\right] + \frac{\mu_S}{\rho} \nabla^2 \vec{v} 
\label{eq:velocityevolutionwdissipation}
\end{equation}
for the velocity field, 
\begin{equation}
\frac{d \rho}{dt} = - \nabla \cdot (\rho \vec{v}) + \mu_{\rho} \nabla^2 \frac{\delta F }{\delta \rho} + \frac{1}{2} \mu_{\rho} \nabla^2 (|\vec{v}|^2)
\label{eq:densityevolutionwdissipation}
\end{equation}
for the density field
and 
\begin{equation}
\frac{d \eta_j }{dt } = - \mathcal{Q}_j \cdot (\eta_j \vec{v}) - \mu_{\eta}  \frac{\delta F }{\delta \eta_j^*}
\label{eq:amplitudeevolutionwithdissipation}
\end{equation}
for the complex amplitudes. Here $\mu_S$, $\mu_{\rho}$ and $\mu_{\eta}$ are dissipation parameters. 

\subsection{Overdamped PFC amplitude expansion}
This is the large $\alpha$ limit of Eq.~\eqref{eq:MPFCAevolution}. The time evolution equations becomes
\begin{equation}
\partial_t \eta_j = - \frac{\delta F}{\delta \eta_j^*}.
\label{eq:overdampedamplitudeevolution}
\end{equation}
Analytically this can be realized by scaling the time as $t \to \alpha t$ in Eq.~\eqref{eq:MPFCAevolution} and taking $\alpha \to \infty$.

\subsection{Overdamped PFC amplitude expansion with mechanical equilibrium}
For this model the dynamics of the system are given by Eq.~\eqref{eq:overdampedamplitudeevolution} with a mechanical equilibrium constraint
\begin{equation}
\frac{\delta F}{\delta \vec{u}}(t) = 0.
\end{equation}
Here $\vec{u}$ is the displacement field that is defined using a decomposition $\eta_j = \phi_j \exp{(i \theta_j ) }$ as 
\begin{equation}
\vec{u} = \frac{2}{3} \sum_{j=1}^3 \vec{q}_j \theta_j.
\end{equation}
The details for this model can be found in Ref.~\cite{Heinonen:2014gr}.

\section{Small displacement limit}
\label{sec:small_displacement_limit}
Here we assume a constant density $\rho=\rho_0$ and complex amplitudes 
\begin{equation}
\eta_j = \phi_0 \exp{[-i \vec{q}_j \cdot \vec{u}(\vec{r},t)]},
\label{eq:small_displacement_amplitude}
\end{equation}
where $\vec{u}$ is a displacement field that is assumed to be small. More specifically the displacement field is expanded up to linear order in the dynamical equations and up to quadratic order in the energy. We also assume that $\vec{u}$ is \emph{slowly varying} and take a long wavelength limit discarding all the derivatives higher than the second order. Expanding the energy up to second order in $\vec{u}$ gives an elastic energy
\begin{equation}
F_{\text{el}} = \int d\vec{r} \left(
\frac{1}{2} \boldsymbol{\sigma} : \boldsymbol{\epsilon}
\right),
\label{eq:elastic_energy}
\end{equation}
where 
\begin{equation}
\boldsymbol{\epsilon}=\frac{1}{2} \left[
\nabla \vec{u} + (\nabla \vec{u})^T
\right]
\end{equation}
is the linear strain tensor and 
\begin{equation}
\boldsymbol{\sigma}=3B^x \phi_0^2 [2 \boldsymbol{\epsilon} + (\nabla \cdot \vec{u}) \mathbf{I}]
\end{equation}
 is the linear elastic stress tensor. The elastic constants can be extracted from a linear relationship $\sigma_{ij} = C_{ijkl} \epsilon_{kl}$ and are those of a 2D hexagonal crystal symmetry. See Appendix \ref{app:elastic_energy} for the derivation of Eq.~\eqref{eq:elastic_energy}.
 
 Some results are needed in order to continue with the analysis. 
 \begin{equation*}
 \begin{split}
 \frac{\delta F}{\delta u_k} &= \sum_j \left(
 \frac{\partial \eta_j^*}{\partial u_k} \frac{\delta F}{\delta \eta_j^*} + \text{C.C.}
 \right) \\
 &= \sum_j \left(
 i q_{j,k} \eta_j^*\frac{\delta F}{\delta \eta_j^*} + \text{C.C.}
 \right) \\ 
 &= 2 \sum_j 
 \Re{\left(i q_{j,k} \eta_j^*\frac{\delta F}{\delta \eta_j^*} \right)} \\ 
 &= - 2 \sum_j 
 \Im{\left(q_{j,k} \eta_j^* \frac{\delta F}{\delta \eta_j^*} \right)} .
 \end{split}
 \end{equation*}
 Here $u_k$ is the $k$th component of $\vec{u}$ and $q_{j,k}$ is the $k$th component of $\vec{q}_j$. Using a different notation we write
 \begin{equation}
  \frac{\delta F}{\delta \vec{u} }  = -2 \sum_j 
 \Im{\left( \vec{q}_j \eta_j^* \frac{\delta F}{\delta \eta_j^*} \right)} .
 \label{eq:functionalddisplacement}
 \end{equation}

We have for all the linear differential operators $\mathcal{L}$ without a constant part a following identity:
\begin{equation}
\mathcal{L} (\phi_0 e^{- i \vec{q}_j \cdot \vec{u}}) 
= - (i \phi_0  \vec{q}_j \cdot \mathcal{L} \vec{u}) e^{- i \vec{q}_j \cdot \vec{u} } + \BigO{|\vec{u}|^2}.
\label{eq:diffidentity}
\end{equation}
Another identity that we need is 
\begin{equation}
\sum_{j=1}^3 \vec{q}_j \otimes \vec{q}_j = \frac{3}{2} \mathbf{I},
\label{eq:reciprocalidentity}
\end{equation}
i.e. the sum of the dyadic of the reciprocal lattice vectors sums up to a constant times identity.  

It should also be pointed out that in the linear regime the functional derivative of the free energy with respect to the displacement field can be approximated as
\begin{equation}
\frac{\delta F}{\delta \vec{u} } \approx \frac{\delta F_{\text{el}}}{\delta \vec{u} } = -3B^x \phi_0^2 \vec{u}^{\sharp},
\label{eq:dfeldu}
\end{equation}
where $\vec{u}^{\sharp}= \nabla^2 \vec{u}+2 \nabla \nabla\cdot \vec{u}$.

\subsection{MPFCA}

Inserting Eq.~\eqref{eq:small_displacement_amplitude} in Eq.~\eqref{eq:MPFCAevolution} gives 
\begin{equation*}
-i \vec{q}_j \cdot (\partial_t^2 \vec{u} + \alpha \partial_t \vec{u} ) \phi_0 e^{-i \vec{q}_j \cdot \vec{u}} = - \frac{\delta F}{\delta \eta_j^*}
\end{equation*}
in the linear displacement regime. Multiplying by $\vec{q}_j \eta_j^*$ on both sides, taking the imaginary part  and summing over $j$ gives
\begin{equation*}
\begin{split}
- \phi_0^2 \left( \sum_{j=1}^3  \vec{q}_j \otimes \vec{q}_j \right) &\cdot  (\partial_t^2 \vec{u} + \alpha \partial_t \vec{u} ) \\
&= - \sum_{j=1}^3 \vec{q}_j \Im{\left(
\eta_j^* \frac{\delta F}{\delta \eta_j^*}
\right)},
\end{split}
\end{equation*}
which becomes 
\begin{equation}
\partial_t^2 \vec{u} + \alpha \partial_t \vec{u} = -\frac{1}{3} \phi_0^{-2} \frac{\delta F}{\delta \vec{u} } 
\end{equation}
with the help of Eq.~\eqref{eq:reciprocalidentity}.

Using Eq.~\eqref{eq:dfeldu} we can write this as 
\begin{equation}
\partial_t^2 \vec{u} + \alpha \partial_t \vec{u} = B^x \vec{u}^{\sharp} . 
\label{eq:MPFCAsmalldisplacement}
\end{equation}
This can be solved with an ansatz $\vec{u} = \vec{u}_0 \exp{(-\omega t + i \vec{k}\cdot \vec{r})}  $. These plane wave solutions can be decomposed into a parallel part $\vec{u}_{\parallel} \exp{(-\omega_{\parallel} t + i \vec{k}\cdot \vec{r})} $, where $\vec{k} \cdot \vec{u}_{\parallel}=ku_\parallel$ and into a perpendicular part $\vec{u}_{\perp} \exp{(-\omega_{\perp} t+i \vec{k}\cdot \vec{r})} $, where $\vec{k} \cdot \vec{u}_{\perp}=0$. The general plane wave solution to Eq.~\eqref{eq:MPFCAsmalldisplacement} is a superposition of these to components with different values of $\vec{k}$. 

Let us solve for the perpendicular mode. Inserting the ansatz into Eq.~\eqref{eq:MPFCAsmalldisplacement} gives
\begin{equation}
\omega_{\perp}^2 - \alpha \omega_{\perp} = - B^x  k^2,
\end{equation}
which can be solved for $\omega_{\perp}$ as
\begin{equation}
\omega_{\perp} = \frac{1}{2} \alpha \pm \frac{i}{2} \sqrt{4 B^x k^2 - \alpha^2},
\label{eq:perpendicular_MPFCA}
\end{equation}
when 
\begin{equation}
k > k_c = \frac{\alpha}{2 \sqrt{B^x}}.
\end{equation}
Here $k_c$ is the critical $k$ value. If this does not hold we get 
\begin{equation}
\omega_{\perp} = \frac{1}{2} \alpha \pm \frac{1}{2} \sqrt{ \alpha^2 - 4 B^x k^2}.
\end{equation}

The calculation for the longitudinal modes gives 
\begin{equation}
\omega_{\parallel} = \frac{1}{2} \alpha \pm \frac{i}{2} \sqrt{12 B^x k^2 - \alpha^2},
\end{equation}
with a critical $k$ value
\begin{equation}
k_c = \frac{\alpha}{2 \sqrt{3 B^x}}.
\end{equation}

\begin{figure}
\includegraphics[width = \columnwidth]{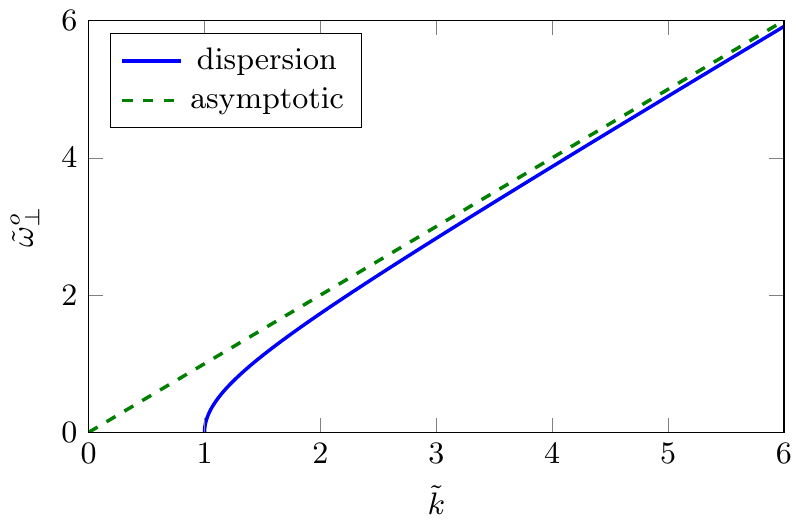}
\caption{The dispersion relation for the oscillating part $\omega_{\perp}^o = \Im{(\omega_{\perp})}$ for the MPFCA model given by Eq.~\eqref{eq:perpendicular_MPFCA}. The units are rescaled as $\tilde{k} = 2 \sqrt{3 B^x } k / \alpha$ and $\tilde{\omega}_{\perp}^o = 2 \omega_{\perp}^o/\alpha$.}
\label{fig:dispersion_MPFCA}
\end{figure}

The dispersion relation for the oscillating part is shown in Fig.~\ref{fig:dispersion_MPFCA}.

\subsection{AMPFCA}

Repeating the previous calculation we get
\begin{equation}
\partial_t^2 \vec{u} + \alpha (-\nabla^2 + 1)\partial_t \vec{u} = B^x \vec{u}^{\sharp},
\label{eq:AMPFCAsmalldisplacement}
\end{equation}
resembling Eq.~\eqref{eq:MPFCAsmalldisplacement}. Here the imaginary part gives a condition $\nabla \cdot \vec{u} = 0$, which is equivalent to the assumption that $\phi_j = \phi_0$ are constant. For details see Appendix \ref{app:AMPFCA_imaginary}.

From hereon the only difference in the calculation is the extra $-\alpha \nabla^2$. This contributes an extra $k^2$ in the final result that can be obtained by replacing $\alpha$ in the previous calculation with $\alpha(1+k^2)$. This gives
\begin{equation}
\omega_{\perp} = \frac{1}{2}\alpha (1+k^2)  \pm \frac{i}{2} \sqrt{ 4 B^x k^2 - (1+k^2)^2 \alpha^2}
\label{eq:perpendicular_AMPFCA}
\end{equation}
for the transversal modes subject to the condition
\begin{equation}
\begin{split}
&\frac{2 B^x}{\alpha^2} - 1  - 2 \frac{\sqrt{B^x}}{\alpha}\sqrt{\frac{B^x}{\alpha^2}-1} 
< k^2 \\
&< \frac{2 B^x}{\alpha^2} - 1  + 2 \frac{\sqrt{B^x}}{\alpha}\sqrt{\frac{B^x}{\alpha^2}-1}.
\end{split}
\end{equation}
For the longitudinal modes we get
\begin{equation}
\omega_{\parallel} = \frac{1}{2}\alpha (1+k^2)  \pm \frac{i}{2} \sqrt{ 12 B^x k^2 - (1+k^2)^2 \alpha^2},
\end{equation}
subject to the condition
\begin{equation}
\begin{split}
&\frac{6B^x}{\alpha^2} - 1  - 2 \frac{\sqrt{3 B^x}}{\alpha}\sqrt{\frac{3 B^x}{\alpha^2}-1} 
< k^2 \\
&< \frac{6B^x}{\alpha^2} - 1  + 2 \frac{\sqrt{3 B^x}}{\alpha}\sqrt{\frac{3 B^x}{\alpha^2}-1}.
\end{split}
\end{equation}

\begin{figure}
\includegraphics[width = \columnwidth]{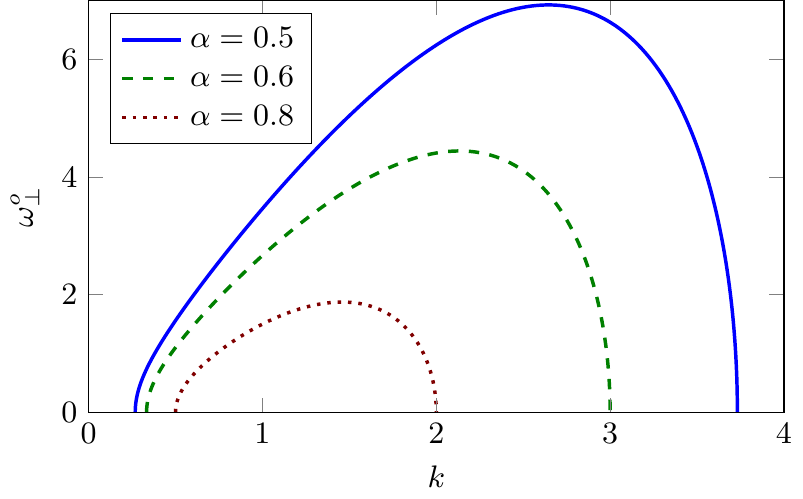}
\caption{The dispersion relation for the oscillating part $\omega_{\perp}^o = \Im{(\omega_{\perp})}$  for the AMPFCA model with several different $\alpha$ given by Eq.~\eqref{eq:perpendicular_AMPFCA}. Here $B^x=1$. }
\label{fig:dispersion_AMPFCA}
\end{figure}

The dispersion relation for the oscillating part is shown in Fig.~\ref{fig:dispersion_AMPFCA}.

\subsection{HPFCA}

Eq.~\eqref{eq:amplitudeevolutionwithdissipation} can be written for the displacement field $\vec{u}$ with the help of Eqs.~\eqref{eq:functionalddisplacement} and \eqref{eq:reciprocalidentity} as
\begin{equation}
\partial_t \vec{u} + \vec{v} \cdot \nabla \vec{u} \approx \partial_t \vec{u} = \vec{v} - \frac{1}{2} \phi_0^{-2} \mu_{\eta} \frac{\delta F_{\text{el}}}{\delta \vec{u}}= \vec{v} +  \mu_{\eta}  B^x \vec{u}^{\sharp}.
\label{eq:smalldisplacementdisplacement}
\end{equation}

The equation for the velocity field, Eq.~\eqref{eq:velocityevolutionwdissipation}, becomes 
\begin{equation}
\begin{split}
\underbrace{\partial_t \vec{v} + \vec{v} \cdot \nabla \vec{v}}_{\approx \partial_t \vec{v}}
&= - \frac{1}{\rho_0} \sum_{j=1}^3 
\left[  \eta_j^* \mathcal{Q}_j 
\frac{\delta F}{\delta \eta_j^*}  
 + \textrm{C.C.}
\right] + \frac{\mu_S}{\rho_0} \nabla^2 \vec{v} \\
& \approx 
- \frac{2}{\rho_0} \sum_{j=1}^3 \Re{
\left[  \eta_j^* i \vec{q}_j
\frac{\delta F}{\delta \eta_j^*}  
\right]} + \mu_S \rho_0^{-1} \nabla^2 \vec{v} \\
&= 
 \frac{2}{\rho_0} \sum_{j=1}^3 \Im{
\left[  \eta_j^* \vec{q}_j
\frac{\delta F}{\delta \eta_j^*}  
\right]} + \mu_S \rho_0^{-1} \nabla^2 \vec{v} \\
& \approx
-\frac{1}{\rho_0} \frac{\delta F_{\text{el}}}{\delta \vec{u}} + \mu_S \rho_0^{-1} \nabla^2 \vec{v} \\
&= 3 B^x \rho_0^{-1} \phi_0^2 \vec{u}^{\sharp} + \mu_S \rho_0^{-1} \nabla^2 \vec{v}.
\end{split}
\label{eq:smalldisplacementvelocity}
\end{equation}
Here  we assume $\nabla (\delta F / \delta \eta_j^*) \approx 0$ since it produces terms with derivatives of the displacement field with a degree higher than two.  

Let $\hat{\xi}$ be the Fourier transform of a vector field $\vec{\xi}$ s.t. $\vec{\xi} = \int d\vec{k} [ \exp{(i \vec{k} \cdot \vec{r})} \hat{\xi}(\vec{k})]$. We can rewrite Eq.~\eqref{eq:smalldisplacementvelocity} as 
\begin{equation}
\partial_t \hat{v} = 3 B^x \rho_0^{-1} \phi_0^2 \hat{u}^{\sharp} - \mu_S \rho_0^{-1} k^2 \hat{v},
\end{equation}
from which we can solve
\begin{equation}
\partial_t (\hat{v} e^{\mu_S \rho_0^{-1} k^2 t}) = 3 B^x \rho_0^{-1} \phi_0^2 \hat{u}^{\sharp}.
\label{eq:smalldisplacementfouriervelocity}
\end{equation}
Now we can write Eq.~\eqref{eq:smalldisplacementdisplacement} in Fourier space as 
\begin{equation}
\partial_t \hat{u} = \hat{v} + \mu_{\eta}  B^x \hat{u}^{\sharp}.
\end{equation}
Multiplying both sides by $\exp{(\mu_S \rho_0^{-1} k^2 t)}$ and taking the time derivative gives
\begin{equation}
\begin{split}
\partial_t [(\partial_t \hat{u} )e^{\mu_S \rho_0^{-1} k^2 t}]  &= \partial_t (\hat{v} e^{\mu_S \rho_0^{-1} k^2 t}) \\
 &+ \mu_{\eta}  B^x \partial_t (\hat{u}^{\sharp} e^{\mu_S \rho_0^{-1} k^2 t}),
 \end{split}
\end{equation}
resulting in 
\begin{equation}
\begin{split}
\partial_t^2 \hat{u} + \mu_S \rho_0^{-1} k^2 \partial_t \hat{u} &= 3 B^x \rho_0^{-1} \phi_0^2 \hat{u}^{\sharp}\\
&+ \mu_{\eta}  B^x ( \partial_t \hat{u}^{\sharp} + \mu_S \rho_0^{-1} k^2\hat{u}^{\sharp} )
\end{split}
\label{eq:fulldispersion}
\end{equation}
with the help of Eq.~\eqref{eq:smalldisplacementfouriervelocity}.
The Fourier transform $\hat{u}^{\sharp}$ is given by
\begin{equation}
\hat{u}^{\sharp} = -k^2 \hat{u} - 2 \vec{k} \vec{k}\cdot \hat{u}.
\end{equation}

\begin{figure}[!h]
\includegraphics[width = \columnwidth]{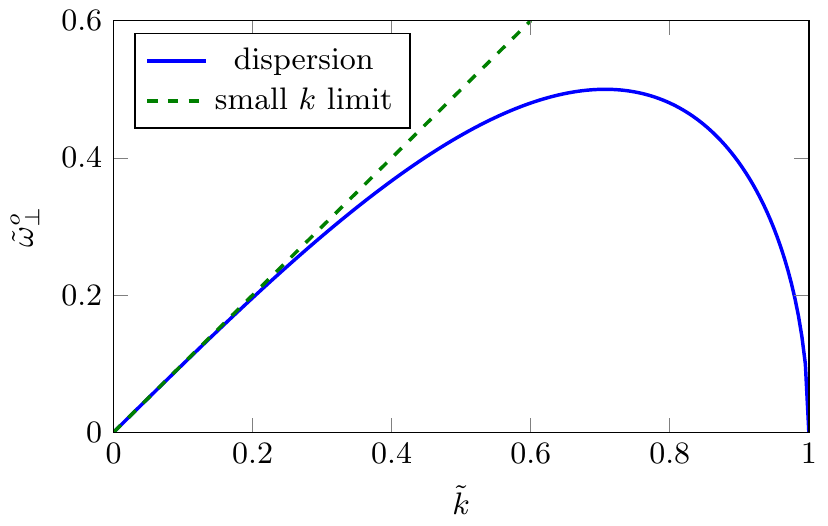}
\caption{The dispersion relation $\tilde{\omega}_{\perp}^o(\tilde{k}) = \tilde{k} \sqrt{1-\tilde{k}^2}$ for the oscillating component of the perpendicular wave in the linear displacement limit for the HPFCA model. 
Here $\tilde{\omega}_{\perp}^o =  |\mu_S  - \frac{3}{2} B^x \mu_{\eta} \rho_0|\omega_{\perp}^o /(6 B^x\phi_0^2 ) $ 
and $\tilde{k}^2 = (\mu_S - \frac{3}{2} B^x \rho_0 \mu_{\eta} )^2  k^2/( 12 B^x\phi_0^2 \rho_0 )$.}
\label{fig:dispersion_hydro}
\end{figure}

Let us split  $\hat{u}$ into two orthogonal parts s.t. $\hat{u} = \hat{u}_{\perp} + \hat{u}_{\parallel}$ and $\vec{k} \cdot \hat{u}_{\perp} = 0$. For transversal modes $\hat{u}^{\sharp} = -k^2 \hat{u}_{\perp}$, giving 
\begin{equation}
\begin{split}
&\partial_t^2 \hat{u}_{\perp} + (\mu_S \rho_0^{-1} +\mu_{\eta}  B^x) k^2 \partial_t \hat{u}_{\perp} \\
 + &3 B^x k^2 (\rho_0^{-1} \phi_0^2 + \frac{1}{3} \mu_{\eta} \mu_S \rho_0^{-1} k^2)
\hat{u}_{\perp} = 0.
\end{split}
\end{equation}
This can be solved with the ansatz $\hat{u}_{\perp} = \exp{(-\omega_{\perp} t)}$ giving
\begin{equation}
\begin{split}
&\omega^2_{\perp} - (\mu_S \rho_0^{-1} +\mu_{\eta}  B^x) k^2 \omega_{\perp}\\
 + &3 B^x k^2 (\rho_0^{-1} \phi_0^2 + \frac{1}{3} \mu_{\eta} \mu_S \rho_0^{-1} k^2) = 0.
 \end{split}
\end{equation}
We solve this for $\omega^2_{\perp} $ resulting in
\begin{equation}
\begin{split}
\omega_{\perp} &= \frac{1}{2} k^2 (\mu_S \rho_0^{-1} +  \mu_{\eta} B^x) \\
&\pm i\frac{k}{2} \sqrt{12 B^x \phi_0^2 \rho_0^{-1} - (\mu_S \rho_0^{-1} - B^x \mu_{\eta})^2 k^2},
\end{split}
\end{equation}
which we can divide into an oscillating part 
\begin{equation}
\omega_{\perp}^o =
\frac{k}{2} \sqrt{12 B^x \phi_0^2 \rho_0^{-1} - (\mu_S \rho_0^{-1} - B^x \mu_{\eta})^2 k^2}
\label{eq:dispersion_hydro_oscillating}
\end{equation}
and a damping part 
\begin{equation}
\omega_{\perp}^d = \frac{1}{2} k^2 (\mu_S \rho_0^{-1} +  \mu_{\eta} B^x)
\end{equation}
s.t. $\omega_{\perp} = \omega_{\perp}^d + i \omega_{\perp}^o$.
The existence of the oscillating solutions is subject to the condition 
\begin{equation}
k < \frac{\sqrt{12 B^x  \rho_0^{-1}}}{\left| \mu_S \rho_0^{-1} - B^x \mu_{\eta} \right|}\phi_0.
\end{equation}
Those modes for which this does not hold are damped with a damping coefficient
\begin{equation}
\begin{split}
\omega_{\perp}^d &= \frac{1}{2} k^2 (\mu_S \rho_0^{-1} +  \mu_{\eta} B^x) \\
&\pm
\frac{k}{2} \sqrt{ (\mu_S \rho_0^{-1} - B^x \mu_{\eta})^2 k^2 - 12 B^x \phi_0^2 \rho_0^{-1} },
\end{split}
\end{equation}
where the sign is determined by the initial velocity. The dispersion relation for the oscillating component is shown in Fig.~\ref{fig:dispersion_hydro}. 

A similar calculation gives an oscillating solution 
\begin{equation}
\begin{split}
\omega_{\parallel} &= \frac{1}{2} k^2 (\mu_S \rho_0^{-1} +  3 \mu_{\eta} B^x) \\
&\pm 
i\frac{k}{2} \sqrt{36 B^x \phi_0^2 \rho_0^{-1} - (\mu_S \rho_0^{-1} - 3 B^x \mu_{\eta})^2 k^2},
\end{split}
\end{equation}
for the longitudinal modes.

\section{Numerical results}
\label{sec:numerical_results}

The numerical results for the AMPFCA model given by the tests presented here were practically indistinguishable from the MPFCA results. For this reason, the AMPFCA results are not presented here separately. The reason for this will be discussed in Sec.~\ref{sec:summary_and_conclusion}.

\subsection{Numerical study of small deformations}

We studied the time evolution of longitudinal waves of the form $\vec{u}(x) =   a \sum_{m=1}^{128} [\sin{(2 \pi m x/L_x)}/m^2]$, where $L_x$ is the size of the periodic box in $x$-dimension and $a$ is the nearest neighbour distance $4 \pi / \sqrt{3}$ \footnote{Any waveform with a large number of modes would do since the modes are not coupled in the linear regime.}. The calculations were performed in a rectangular box with periodic boundary conditions. The dimensions of the box were $(L_x,L_y)=(8192,256)$. Note that the displacement field varies only in one direction reducing the system to one dimension. Here we use a parametrization $\alpha=0.05$, $\mu_S=0$, $\mu_{\rho} = 0.05$, $\mu_{\eta} = 1$, $B^x=\tilde{B}^x=1$, $\Delta B = 0.097$, $\tau=0.885$, $\nu = 1$ and the average density $\rho_0=0.1$. For the numerical discretization we used $\Delta x=\Delta y=4$ and the time step $\Delta t$ was varied from 0.0625 to 0.125. For more on numerical details see Appendix \ref{app:numerical_methods}.

\begin{figure}
\includegraphics[width = \columnwidth]{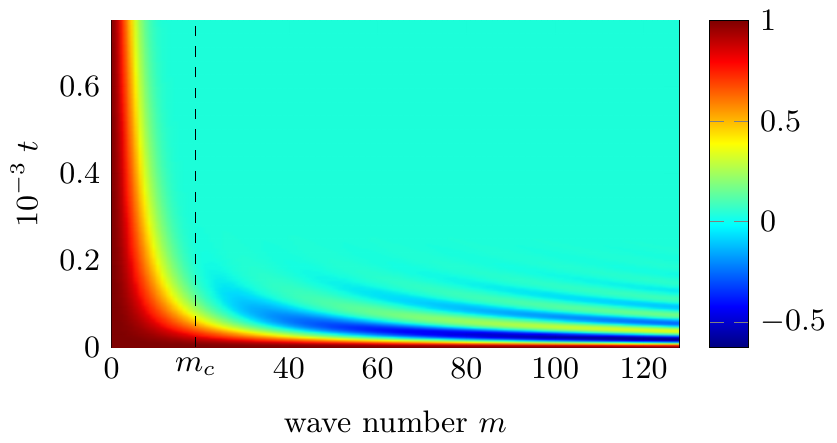}
\caption{The spectrum of the MPFCA model as a function of time. The color shows the relative amplitude of the $m$th mode at a given time. The amplitudes are scaled to unity at time zero. The wave number $m$ denotes the $m$th harmonic. The wave vectors can be recovered as $k_m = 2\pi m /L_x$. The analytical cutoff for the oscillating solutions is shown here at wave number $m_c$. }
\label{fig:specturm_MPFCA}
\end{figure}
\begin{figure}
\includegraphics[width = \columnwidth]{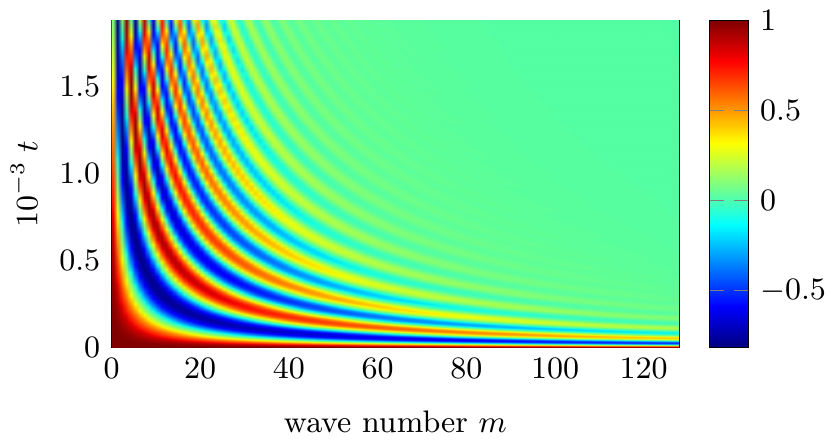}
\caption{The spectrum of the HPFCA model as a function of time. The color shows the relative amplitude of the $m$th mode at a given time. The amplitudes are scaled to unity at zero. The wave number $m$ denotes the $m$th harmonic. The wave vectors can be recovered as $k_m = 2\pi m /L_x$. The time difference between the peaks scales with the wave number as $m^{-2}$ since the dissipation rate is proportional to $k^2$.}
\label{fig:specturm_HPFCA}
\end{figure}

The results are presented in Figs.~\ref{fig:specturm_MPFCA} and \ref{fig:specturm_HPFCA}. For the MPFCA model there is a critical wave vector $k_c$ below which no oscillating solutions exists. It can also be seen that the oscillating solutions above $k_c$ are damped at a rate that is independent of $k$. The HPFCA model behaves very differently showing oscillating solutions for all $k$ and damping dependent on $k$. 

\subsection{Grain rotation}

In order to compare the different dynamics, we solve the time evolution of a circular grain embedded in a crystalline matrix. This numerical test is simple but nontrivial and should provide insight into the difference of the dynamical schemes discussed in this article. 

\begin{figure}
\subfigure[]
{\includegraphics[width=4cm]{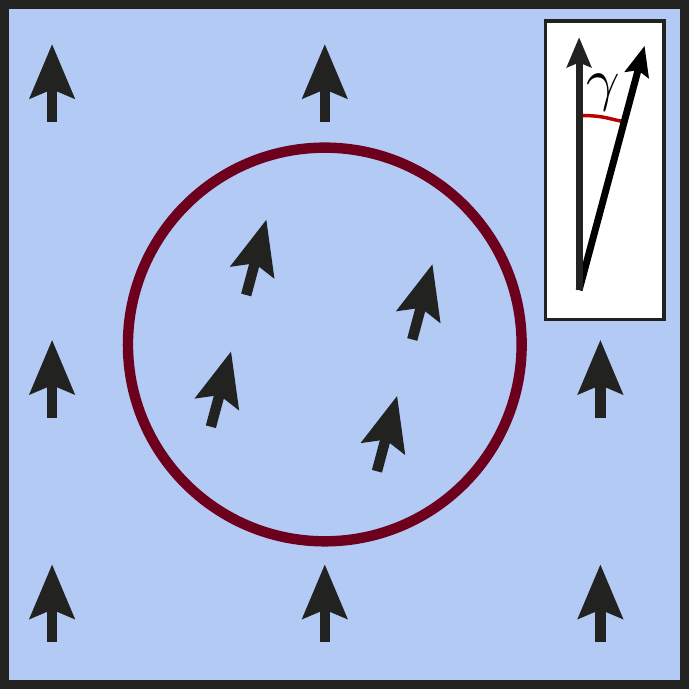}}\;
\subfigure[]
{\includegraphics[width=4cm]{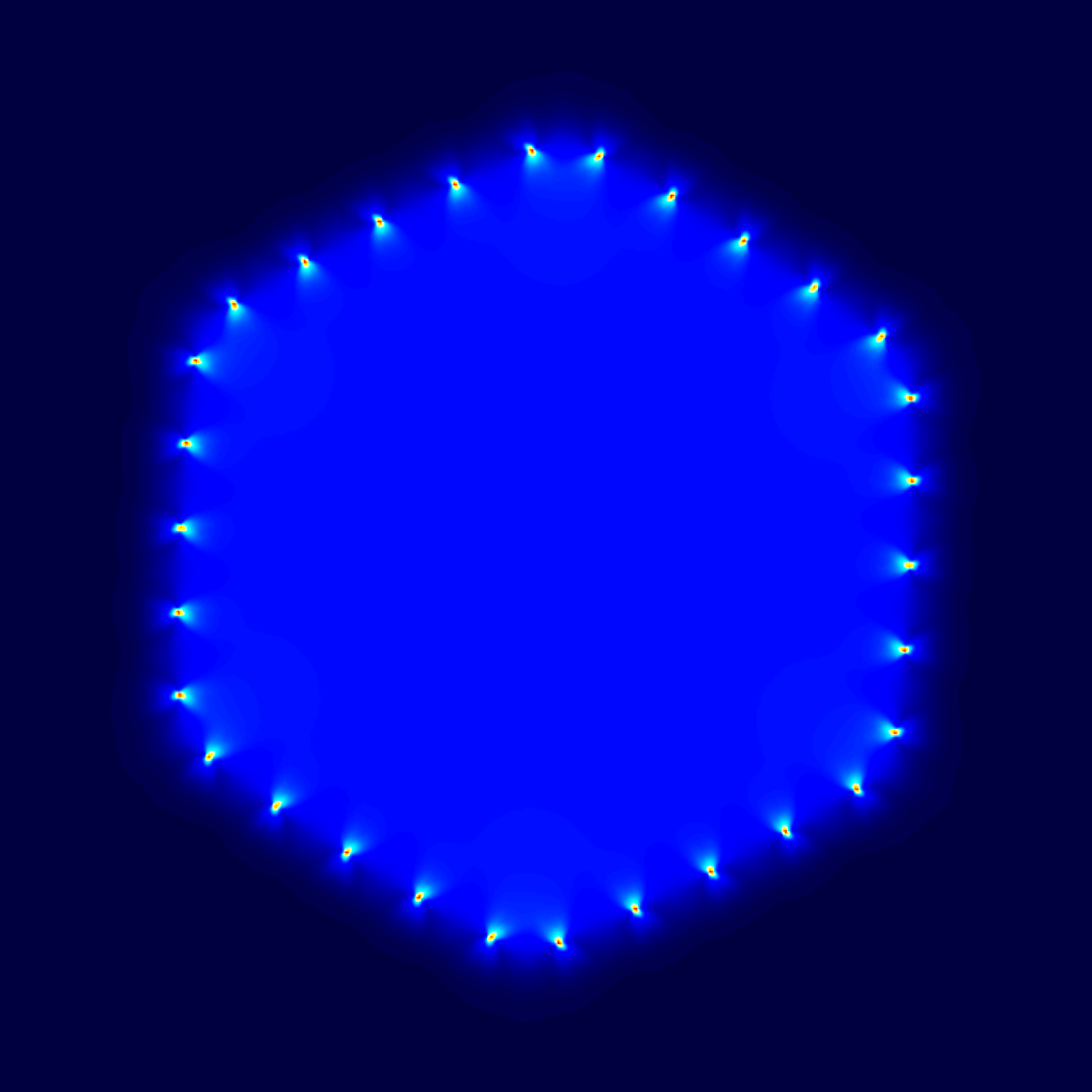}}
\caption{(a) a schematic of the grain rotation calculation: the grain is rotated by an angle $\gamma$ (the angle difference is shown in the inset). (b) The magnitude of the gradient of the displacement field $|\nabla \vec{u}|$ from a calculation showing the dislocation cores at the perimeter of the grain. The uniform brighter color inside the grain is due to the angle difference $\gamma$.
}
\label{fig:grain}
\end{figure}

The grain is tilted by an angle $\gamma(t=0)$ creating a mismatch at the perimeter of the grain, which gives rise to a grain boundary. The setting is shown in Fig.~\ref{fig:grain}. 

Assuming that the grain boundary motion is curvature driven i.e. $v_{\perp} = \partial_t R \sim \kappa = R^{-1}$, we can solve for the time evolution of the radius $R(t)$ giving $\partial_t (R^2) = \text{constant}$. This implies that the area of the grain decreases linearly. Note that if the normal velocity of the grain boundary is proportional to the curvature of the grain boundary, the time evolution of a circular grain is self similar in the sense that the grain will be circular also at later times. 

At small angles the number of the dislocation cores $n_d$ at the perimeter of the rotated grain is proportional to the rotation angle and the length of the boundary. This can be written as $n_d \sim \gamma(t) R(t)$. Dislocation cores repel each other and annihilate in the very end of the calculation. For earlier times $n_d$ is constant in time implying that $\gamma(t) \sim R^{-1}$. From this it follows that angle $\gamma$ increases as radius $R$ decreases. The dynamics of this type of rotated grain is discussed in depth in Ref.~\cite{Adland:2013ew}.

\begin{figure}
\includegraphics[width = \columnwidth]{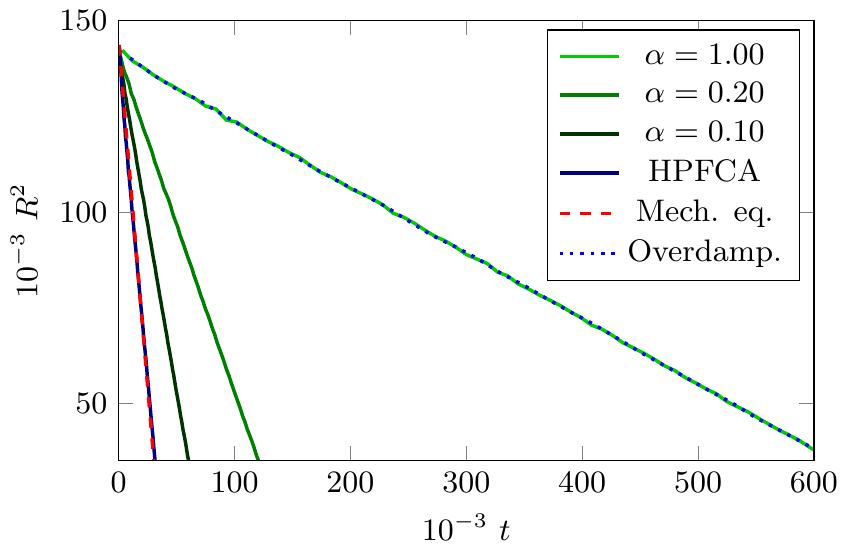}
\caption{The squared radii of the shrinking circular grain with different models and parametrization. The overdamped model and MPFCA with $\alpha=1$ give the slowest dynamics. MPFCA dynamics becomes faster with decreasing $\alpha$ but cannot reach the fastest trajectories given by the mechanically equilibrated model and the HPFCA model. }
\label{fig:radii}
\end{figure}

The results of the grain rotation calculations are shown in Fig.~\ref{fig:radii}. All the different realizations show linear time evolution for the area of the rotated grain. The behaviour of the MPFCA model reduces to the overdamped amplitude model when $\alpha=1$. More interestingly, the trajectory for the MPFCA model converges to the one obtained with $\alpha=0.1$. Decreasing $\alpha$ further did not make the dynamics faster. The HPFCA model gives the same trajectory as the mechanically equilibrated overdamped model as already discovered in Ref.~\cite{Heinonen:2015}. 

The parameters used for these calculations were $\mu_S=0$, $\mu_{\rho} = 0.05$, $\mu_{\eta} = 1$, $B^x=\tilde{B}^x=1$, $\Delta B = 0.097$, $\tau=0.885$, $\nu = 1$ and the average density $\rho_0=0.1$. 

The calculations were performed in a box of a size $L_x = L_y = 1536$ with a discretization $\Delta x=\Delta dy=2$, $\Delta t=0.125$. For more details see Appendix \ref{app:numerical_methods}.

\section{Summary and conclusion}
\label{sec:summary_and_conclusion}

We have analysed three different schemes for the time evolution of the PFC amplitude system analytically and numerically. We have shown that no true phonons exist for other models than the  HPFCA model. The analytical results for the small displacements are verified numerically showing that the damping of the oscillating solutions for the MPFCA model dissipate at the same rate regardless of the wavelength and that there is a critical wavelength over which the waves do not oscillate. All the different modes of the HPFCA model oscillate and the dissipation is proportional to $k^2$.

The grain rotation experiment shows that in case of the MPFCA model, there is a critical value for the dissipation parameter $\alpha$ below which the dynamics does not get faster. Unlike for the HPFCA model, the limiting trajectory is not that of the mechanically equilibrated system. Instead, slower time evolution is seen. This behaviour should not be caused by the cutoff in the oscillating solutions  since $k_c$ can be controlled by decreasing $\alpha$ allowing all the modes in the periodic box to oscillate. Instead, it is likely that this behaviour follows from the fact that the parameter $\alpha$ controls all the dissipation in the system. Even if the oscillating modes are damped less allowing for reducing the energy through the displacement field, the overall dissipation is reduced hindering the diffusional relaxation of the system. Taking $\mu_{S}$ to 0 for the HPFCA model does not affect the diffusional dissipation that happens through the parameter $\mu_{\eta}$. 

The AMPFCA results are not shown in the numerical experiments since they are indistinguishable from the results given by the MPFCA model. It seems that at large $\alpha$ both models collapse into the overdamped case and at low $\alpha$ they become the same. Studying the small displacement dispersion relations of these two models can give some insight into why this happens. Let us consider Eqs.~\eqref{eq:perpendicular_MPFCA} and \eqref{eq:perpendicular_AMPFCA}. Expanding the dispersion relation up to a quadratic order in both $\alpha$ and $k$ Eq.~\eqref{eq:perpendicular_AMPFCA} becomes
\begin{equation*}
\begin{split}
\omega_{\perp} &= \frac{1}{2} \alpha + \BigO{\alpha k^2} \pm \frac{i}{2} \sqrt{4 B^x k^2 - \alpha^2 + \BigO{\alpha^2 k^2}} \\
& \approx \frac{1}{2} \alpha \pm \frac{i}{2} \sqrt{4 B^x k^2 - \alpha^2 },
\end{split}
\end{equation*}
giving the dispersion relation for the MPFCA model i.e. Eq.~\eqref{eq:perpendicular_MPFCA}. The expansion in $\alpha$ is justified since $\alpha$ has to be small in order to have high wavelength oscillating solutions. The amplitude energy $F$ penalises high $k$ modes and they are rarely seen in the calculations. Modes with $k=1$ correspond to oscillations at interatomic distance implying that it is a relatively good approximation to state that for the large scale displacements $k \ll 1$ . 

The results presented here suggest that while the MPFCA model remains a good qualitative description of fast dynamics, it is not suitable for separating the time scales of the lattice vibrations and the diffusional phenomena. The relaxation times of grain boundaries and slow phenomena are always coupled through parameter $\alpha$ to the relaxation of elastic excitations. This should become more important when the system size is increased. 

\begin{acknowledgments}
This work has been supported in part by the Academy of Finland through its COMP CoE Grants No. 251748 and 284621. The authors wish to acknowledge CSC IT Center for Science, Finland, for generous computational resources. We acknowledge the computational resources provided by the Aalto Science-IT project.
\end{acknowledgments}

\appendix

\section{Derivation of Eq.~(\ref{eq:elastic_energy}) }
\label{app:elastic_energy}
Here we expand the free energy $F$ to a quadratic order in the displacement field $\vec{u}$. We write the complex amplitudes as $\eta_j = \phi_0 \exp{( -i\vec{q}_j \cdot \vec{u} )}$. We assume that the amplitudes of the complex fields are constant. Looking at Eq.~\eqref{eq:energy} we see that only parts where the phase of the complex fields matter are the parts with the operator $\mathcal{G}_j$ and the terms $\eta_1 \eta_2 \eta_3$ and its complex conjugate. The latter terms give 
\begin{equation}
\prod_{j=1}^3 \eta_j = \phi_0^3 e^{ -i \vec{u} \cdot \sum_{j=1}^3 \vec{q}_j} = \phi_0^3,
\end{equation}
since $\sum_{j=1}^3 \vec{q}_j = 0$. 

The only remaining part is
\begin{equation}
f_{\text{el}} := B^x \sum_{j=1}^3 \left| \mathcal{G}_j \phi_0 e^{ -i\vec{q}_j \cdot \vec{u} } \right|^2.
\end{equation}
Using the identity given by Eq.~\eqref{eq:diffidentity} we get
\begin{equation}
\begin{split}
f_{\text{el}} &= B^x \phi_0^2 \sum_{j=1}^3 \left| -i \vec{q}_j \cdot \mathcal{G}_j \vec{u} + \BigO{|\vec{u}|^2}  \right|^2 \\
&= B^x \phi_0^2 \sum_{j=1}^3 \left| 
-i \vec{q}_j \cdot \nabla^2 \vec{u} + 2 \vec{q}_j \cdot (\vec{q}_j \cdot \nabla) \vec{u} + \BigO{|\vec{u}|^2 }
\right|^2 \\
&= B^x \phi_0^2 \sum_{j=1}^3 \left\lbrace
\left( \vec{q}_j \cdot \nabla^2 \vec{u} \right)^2 \right. \\ & \left. 
+ 4 \left[ \vec{q}_j \cdot (\vec{q}_j \cdot \nabla) \vec{u}  \right ]^2 + \BigO{|\vec{u}|^3 }
\right\rbrace.
\end{split}
\end{equation}
We take the long wavelength limit discarding higher order derivatives and the terms of order higher than two and get
\begin{equation}
f_{\text{el}} \approx 4 B^x \phi_0^2 
\sum_{j=1}^3 \left[ \vec{q}_j \cdot (\vec{q}_j \cdot \nabla) \vec{u} \right]^2.
\end{equation}
Here it is convenient to use the Einstein summation convention: now
\begin{equation}
f_{\text{el}} \approx 4 B^x \phi_0^2 q_{j,i_1} q_{j,i_2} q_{j,i_3} q_{j,i_4}  (\partial_{i_1} u_{i_2}) (\partial_{i_3} u_{i_4}).
\end{equation}
Here $q_{j,i_1}$ denotes the component $i_1$ of the vector $\vec{q}_j$. It can be shown that 
\begin{equation}
\begin{split}
q_{j,i_1} q_{j,i_2} q_{j,i_3} q_{j,i_4} &= \frac{3}{8} ( \delta_{i_1, i_2} \delta_{i_3, i_4}  \\
&+ \delta_{i_1, i_3} \delta_{i_2, i_4} + \delta_{i_1, i_4} \delta_{i_2, i_3} )
\end{split}
\end{equation}
by using Eq.~\eqref{eq:reciprocalidentity} and the fact that $\sum_j \vec{q}_j = 0$ (or in coordinates with a lengthy calculation). Now 
\begin{equation}
\begin{split}
f_{\text{el}} &\approx  B^x \phi_0^2 \frac{3}{2} [
(\partial_{i_1} u_{i_1}) (\partial_{i_3} u_{i_3})  \\
& +(\partial_{i_1} u_{i_2}) (\partial_{i_1} u_{i_2} ) + (\partial_{i_1} u_{i_2}) (\partial_{i_2} u_{i_1} )].
\end{split}
\end{equation}
Using the definition $\epsilon_{ij} = \frac{1}{2} (\partial_i u_j + \partial_j u_i)$ we can recast this in a form 
\begin{equation}
f_{\text{el}}  \approx \frac{3}{2} B^x \phi_0^2 \left(
\partial_k u_k \delta_{ij} + 2 \epsilon_{ij}
\right) \epsilon_{ij} ,
\end{equation}
from which we recover Eq.~\eqref{eq:elastic_energy} using the definition $\boldsymbol{\sigma} : \boldsymbol{\epsilon} = \sigma_{ij} \epsilon_{ij}$.

\section{The imaginary part of the small displacement equation for the AMPFCA model}
\label{app:AMPFCA_imaginary}

Let us assume at first that $\eta_j(\vec{r},t) = \phi_j(\vec{r},t) \exp{[i \theta_j(\vec{r},t)]}$. Now 
\begin{equation}
\begin{split}
\phi_j \frac{\delta F}{\delta \phi_j} & = \phi_j \frac{\partial \eta_j^*}{\partial \phi_j} \frac{\delta F}{\delta \eta_j^*} + \text{C.C.} \\
&= \phi_j e^{-i \theta_j} \frac{\delta F}{\delta \eta_j^*} + \text{C.C.} \\
&= \eta_j^* \frac{\delta F}{\delta \eta_j^*} + \text{C.C.} \\
& = 2 \Re{ \left(
\eta_j^* \frac{\delta F}{\delta \eta_j^*} 
\right) }.
\end{split}
\label{eq:amplitudefunctionalderivative}
\end{equation}

Next we assume that the displacements are small and that $\eta_j = \phi_0 \exp{(-i \vec{q}_j \cdot \vec{u})}$. Inserting Eq.~\eqref{eq:small_displacement_amplitude} in Eq.~\eqref{eq:AMPFCAevolution} gives
\begin{equation}
- i\vec{q}_j \cdot (\partial_t^2 \vec{u} - \alpha \mathcal{Q}_j^2 \partial_t \vec{u}) \phi_0 e^{-i \vec{q}_j \cdot \vec{u}}
= -\frac{\delta F}{\delta \eta_j},
\end{equation}
with the help of Eq.~\eqref{eq:diffidentity}. Multiplying by $\vec{q}_j \eta_j^*$, taking the imaginary part and summing over $j$ gives Eq.~\eqref{eq:AMPFCAsmalldisplacement} but here we look at the real part. Multiplying by $\eta_j^*$ and taking the real part gives
\begin{equation}
2\phi_0^2 (\vec{q}_j \otimes \vec{q}_j) : \nabla \vec{u} = \frac{1}{2} \phi_0 \left. \frac{\delta F}{\delta \phi_j}\right|_{\phi_j=\phi_0}
\label{eq:app_displacement_eq}
\end{equation}
using Eq.~\eqref{eq:amplitudefunctionalderivative}.

Assuming that $\phi_j$ is constant implies that the chemical potential $\delta F / \delta \phi_j = 0$. Otherwise $\phi_j$ would change in time. This implies that the right hand side of Eq.~\eqref{eq:app_displacement_eq} is 0. Summing over $j$ gives
\begin{equation}
\mathbf{I} : \nabla \vec{u} = \nabla \cdot \vec{u} = 0.
\end{equation}
The assumption that $\phi_j=\phi_0$ implies that the system is incompressible. Other way to see that $\delta F / \delta \phi_j = 0$ is to calculate 
\begin{equation}
\frac{\delta F}{\delta \phi_0 } = \frac{\delta F_{\text{el}}}{\delta \phi_0} 
= 3 B^x \phi_0 [2 \boldsymbol{\epsilon}: \boldsymbol{\epsilon} + (\nabla \cdot \vec{u})^2 ] = \BigO{|\nabla \vec{u}|^2}.
\end{equation}
This gives 0 since in the dynamical equations we assume that $\vec{u}$ is of linear order.

\section{Numerical methods}
\label{app:numerical_methods}
All the calculations were solved using a semi-implicit algorithm \cite{Tegze2009,Vollmayr-Lee2003,Zhu1999}, where the linear terms of the form $\mathcal{L}(\nabla) \psi(\vec{r}) $ are treated implicitly, while the non-linear parts are treated explicitly. All the derivatives were computed in $k$-space. 

\bibliography{fast_times_comparison}

\end{document}